\documentclass{article}
\usepackage{src/ijcai25}

\usepackage{times}
\usepackage{soul}
\usepackage{url}
\usepackage[hidelinks]{hyperref}
\usepackage[utf8]{inputenc}
\usepackage[small]{caption}
\usepackage{graphicx}
\usepackage{amsmath}
\usepackage{amsthm}
\usepackage{booktabs}
\usepackage{algorithm}
\usepackage{algorithmic}
\usepackage[switch]{lineno}

\usepackage{bm}
\usepackage{xcolor}
\usepackage{enumitem}
\usepackage{amssymb}
\usepackage{float}
\usepackage{microtype}

\newcommand{\citet}[1]{\citeauthor{#1} \shortcite{#1}}

\hypersetup{
    colorlinks=true,
    linkcolor=blue,
    citecolor=blue,
    filecolor=magenta,      
    urlcolor=cyan,
    pdftitle={Overleaf Example},
    pdfpagemode=FullScreen,
}

\title{AudioMorphix: Training-free audio editing with diffusion probabilistic models}

\author{
Jinhua Liang$^1$\and
Yuanzhe Chen$^{3}$\and
Yi Yuan$^2$\and
Dongya Jia$^{3}$\and\\
Xiaobin Zhuang$^{3}$\and
Zhuo Chen$^{3}$\and 
Yuping Wang$^{3}$\and
Yuxuan Wang$^{3}$\\
\affiliations
$^1$Queen Mary University of London\\
$^2$University of Surrey\\
$^3$ByteDance\\
\emails
jinhua.liang@qmul.ac.uk
}
\begin{document}

\maketitle

\begin{figure*}[!ht]
    \centering
    \includegraphics[width=0.62\linewidth]{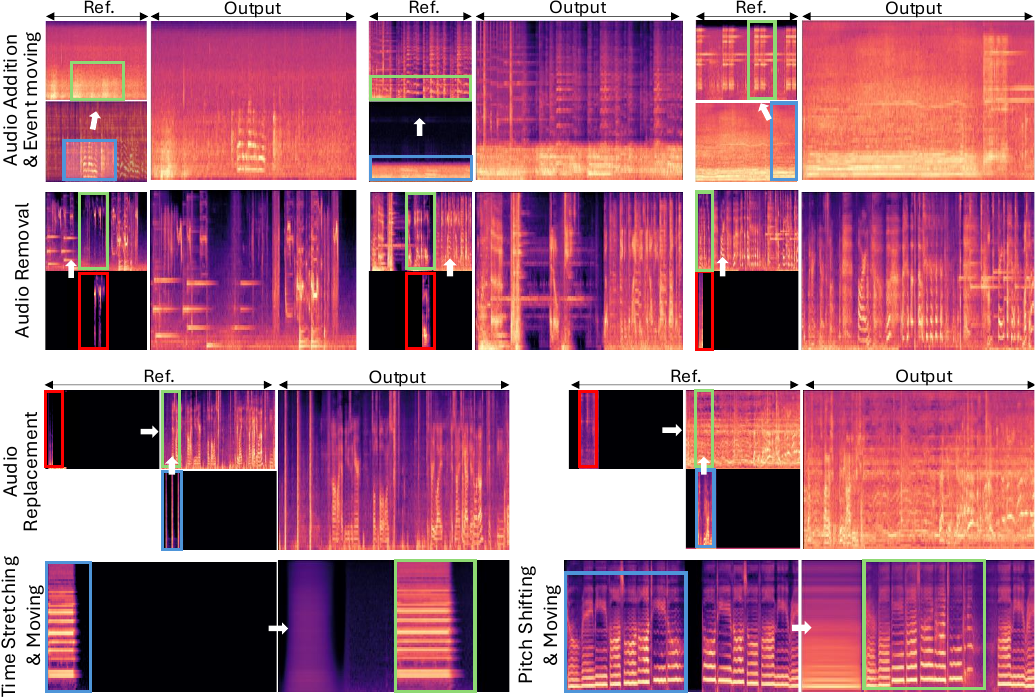}
    \caption{Audio editing tasks of which our AudioMorphix is capable with no training cost. We use \textcolor{green}{green} to highlight the editing region on the source audio while \textcolor{blue}{blue} and \textcolor{red}{red} indicate the regions for addition and removal, respectively. The arrows represent the direction of the editing process, showing the flow from the reference audio to the source audio.}
    \label{fig:demo}
    \vspace{-0.3cm}
\end{figure*}

\begin{abstract}
Editing sound with precision is a crucial yet underexplored challenge in audio content creation. While existing works can manipulate sounds by text instructions or audio exemplar pairs, they often struggled to modify audio content precisely while preserving fidelity to the original recording. In this work, we introduce a novel editing approach that enables localized modifications to specific time-frequency regions while keeping the remaining of the audio intact by operating on spectrograms directly. To achieve this, we propose \textbf{AudioMorphix}, a training-free audio editor that manipulates a target region on the spectrogram by referring to another recording. Inspired by morphing theory, we conceptualize audio mixing as a process where different sounds blend seamlessly through morphing and can be decomposed back into individual components via demorphing. Our AudioMorphix optimizes the noised latent conditioned on raw input and reference audio while rectifying the guided diffusion process through a series of energy functions. Additionally, we enhance self-attention layers with a cache mechanism to preserve detailed characteristics from the original recordings. To advance audio editing research, we devise a new evaluation benchmark, which includes a curated dataset with a variety of editing instructions. Extensive experiments demonstrate that AudioMorphix yields promising performance on various audio editing tasks, including addition, removal, time shifting and stretching, and pitch shifting, achieving high fidelity and precision. Demo and code are available at \href{https://jinhualiang.github.io/AudioMorphix-Demo/}{this url}.
\end{abstract}

\section{Introduction} \label{sec:introduction}
Generative modeling~\cite{ho_denoising_2020,song_denoising_2022,liang2025acoustic} has witnessed rapid breakthroughs in recent years, particularly in the domain of denoising diffusion models. Despite progress was primarily seen in image synthesis, applications of audio generation have been attracting increasing interests~\cite{liu_audioldm_2023,ghosal2023tango,majumder2024tango,liu_wavjourney_2023}. While audio diffusion methods~\cite{liu_audioldm_2023,huang_make--audio_2023} are capable of generating diverse, high-fidelity audio, designing plausible guidance signals to create content consistent with user preference remains challenging.

Recent works have explored manipulating the audio content to align with user preference better. \citet{wang_audit_2023} trained an end-to-end latent diffusion model to edit the audio content using text instructions. \citet{manor_zero-shot_2024} and \citet{liu_audioldm_2023} achieved editing by inverting raw audio into noisy latent and re-sample sounds from the obtained latent with new instructions. \citet{cheng2025audiotexturemanipulationexemplarbased} modified audio recordings by learning from paired audio exemplars. Nevertheless, it is still challenging for these editing methods to precisely change the audio content within a specific region while keeping the remaining of the audio intact, limiting their ability to fully meet user preference. Thus, we introduce a novel editing approach that facilitates editing on local time-frequency (T-F) regions while maintaining the rest unchanged by operating on spectrograms directly. Compared to previous editing methods, the advantages of the spectrogram-based approach are three-fold: (1) Preciseness: Direct operation on the spectrogram disentangles the sound into frequency bins, allowing the manipulation of partial sound components without affecting the remaining elements. (2) Fidelity: This approach preserves the uninterested regions, ensuring they are unchanged throughout the editing process. (3) Interactivity: Users can change the audio content by drawing a mask over the region of interest and providing reference audio, enabling an intuitive editing process.

In this paper, we propose AudioMorphix, a novel training-free sound editor that manipulates raw audio recordings conditioned on reference audio and a binary spectrogram mask. In particular, we cast audio editing as part of a \textit{morphing cycle} performed on the latent space manifold \cite{he_manifold_2023,yang2024impus}: a sound mixture is created by morphing different recordings, while individual tracks are separated by demorphing the mixtures. Consequently, common audio editing tasks like addition and removal can be specified as a latent morphing traversal on the manifold. We design various energy functions to guide the sampling process of latent diffusion models~\cite{liu_audioldm_2023,ghosal2023tango,majumder2024tango}. Additionally, we preserve the details of raw audio by substituting key and value components of self-attention layers in the current diffusion process with those from the reference audio, following empirical findings in the image domain. To evaluate a broader range of audio editing methods, as shown in Figure~\ref{fig:demo}, we collect a new audio editing dataset that enables manipulation of raw audio content using various prompts, such as paired text description, task instruction, and reference audio.

Experiment results show that the proposed AudioMorphix outperforms state-of-the-art audio editing models on audio addition, removal, replacement, moving, pitch shifting, and time stretching tasks. We also examine the impact of various system factors by ablation study of our proposed components.

In summary, the contributions of this paper are as follows:
\begin{itemize}[itemsep=1pt, topsep=1pt, partopsep=1pt]
    \item We introduce a new editing approach in which a target time-frequency region is manipulated on spectrograms directly. Compared to existing editing methods, this spectrogram-based approach enables precise audio modification while preserving the rest of the regions unchanged.
    \item We propose a training-free framework that edits the specific region of raw audio by using pre-trained audio diffusion models. We devise energy functions to guide audio generation along the trajectory of the diffusion process. We also manipulate the features in self-attention layers by substituting the key and value components of the generation with those from the reference sounds.
    \item We create a new dataset to compare a range of audio editing methods by assessing the generated audio on the particular T-F area. We show AudioMorphix outperforms current state-of-the-art methods on various tasks, including addition, removal, replacement, time shifting and stretching, and pitch shifting.
\end{itemize}

\begin{figure*}[!ht]
    \centering
    \includegraphics[width=0.7\linewidth]{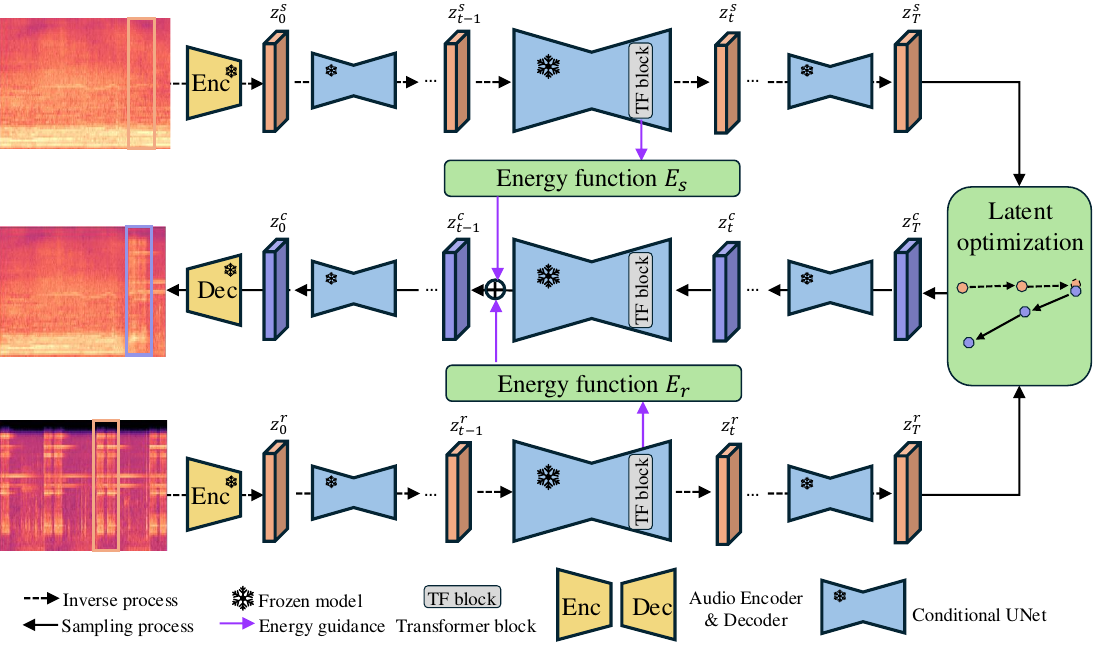}
    \vspace{-0.1cm}
    \caption{Overview of the proposed AudioMorphix. The AudioMorphix generates the clean latent $z_0^c$ by rectifying the sampling process with latent optimization and energy guidance: After obtaining the noisy latents  $z_T^s$  and  $z_T^r$  from the spectrograms of the raw and reference audio, along with their corresponding text descriptions $s$ and $r$, AudioMorphix updates the noisy estimate $z_T^c$, where $c$ represents the target text description, during the morphing cycle (in Section~\ref{subsec:manipulate_latent_in_morphing_cycle}); Throughout the sampling process, AudioMorphix estimates the latent $z_{t-1}^{c}$ via a frozen latent diffusion model, guided by energy-based functions (in Section~\ref{subsec:guide_audio_editing_with_energy_function}).}
    \label{fig:overview}
    \vspace{-0.3cm}
\end{figure*}
    
\section{Preliminaries} \label{sec:preliminaries}
\subsection{Denoising Diffusion Models} \label{subsec:diffusion_models}
\textbf{Denoising diffusion models}. Diffusion models, or score-matching networks, have achieved great process in high-quality generation across various domains, such as image~\cite{Dhariwal2021,Zhang2022Inversion-BasedModels}, video~\cite{Xie2024SV4DD3}, symbolic music~\cite{Zhang2024DExterModels} and audio generation~\cite{liu_audioldm_2023}. Let $\mathbf{x}\in \textnormal{X}\subset \mathbb{R}^d$ be a $d$-dimentional sample in the finite set of $\textnormal{X}$, drawn from the ``true but unknown'' distribution $P$, and $\mathbf{y}\in \textnormal{Y}$ be the provided condition, such as text description. Diffusion models generate a new sample by a sequence of invocation of time-dependent score function $\nabla_{\mathbf{x}_t} \log p_t(\mathbf{x}_t)$ for noisy data $\mathbf{x}_t$. During training, a noise variable $\boldsymbol{\epsilon}$ is sampled from Gaussian distribution $\boldsymbol{\epsilon}\sim\mathcal{N}(\mathbf{0},\bm{I})$. The noisy data $\mathbf{x}_t$ is obtained as a linear combination of the noise variable $\boldsymbol{\epsilon}$ and the clean data $\mathbf{x}_0\sim P(\mathbf{x})$ at the step $t$, as $\mathbf{x}_t = \sqrt{\overline{\alpha}_t} \mathbf{x}_0 + \sqrt{1 - \overline{\alpha}_t} \boldsymbol{\epsilon}$ where $\overline{\alpha}_t > 0$ is a scaling parameter. This conditional probability distribution can be defined by $q(\mathbf{x}_t|\mathbf{x}_0):=\mathcal{N}(\mathbf{x}_t; \sqrt{\overline{\alpha}_t} \mathbf{x}_0, (1-\overline{\alpha}_t){\bm{I}})$. A diffusion model learns a denoiser $\boldsymbol{\epsilon}_{\theta}(\mathbf{x}_t, t)$ to parameterize the score function with the loss function
\begin{equation}
   \mathbb{E}_{\mathbf{x}_0, t, \boldsymbol{\epsilon}_t \sim \mathcal{N}(0, 1)} \left[ \left\| \boldsymbol{\epsilon}_t - \boldsymbol{\epsilon}_{\theta}(\mathbf{x}_t, t) \right\|_2^2 \right], 
\end{equation}
where $\theta$ is a set of learnable parameters of the denoiser. In the sampling process, we apply the denoiser $\boldsymbol{\epsilon}_{\theta}$ to estimate the noise variable $\boldsymbol{\epsilon}_{t-1}$ and substitute it from noisy data $\mathbf{x}_t$ iteratively to get the clean data $\mathbf{x}_0$.

\textbf{Denoising diffusion implicit models (DDIM)}. DDIM was proposed to improve the inference speed by using a deterministic generative process~\cite{ho_denoising_2020}. During inference, DDIM obtains noisy data $x_{t-1}$ at the step $t$ with the following update rule:
\begin{align}
\mathbf{x}_{t-1} &= \sqrt{\bar{\alpha}_{t-1}} \left( \frac{\mathbf{x}_t - \sqrt{1 - \bar{\alpha}_t} \boldsymbol{\epsilon}_{\theta}(\mathbf{x}_t, t)}{\sqrt{\bar{\alpha}_t}} \right)\nonumber \\
&+ \sqrt{1 - \bar{\alpha}_{t-1} - \sigma_t^2}\boldsymbol{\epsilon}_{\theta}(\mathbf{x}_t, t) + \sigma_t \boldsymbol{\epsilon}_t,
\end{align}
where on the right side the first term is a prediction of the clean data $\mathbf{x}_0$ using the noisy data $\mathbf{x}_t$ and the denoiser $\boldsymbol{\epsilon}_{\theta}$, the second term represents the estimated direction pointing to $\mathbf{x}_t$, and the last term denotes a random noise. $\sigma_t$ is a scaling factor controlling the stochasticity in the sampling process: with $\sigma_t = \sqrt{(1 - \bar{\alpha}_{t-1}) / (1 - \bar{\alpha}_t)} \sqrt{1 - \bar{\alpha}_t / \bar{\alpha}_{t-1}}$. DDIM is implemented as DDPM while $\sigma_t =0$ is interpreted as a deterministic sampling process. It is noteworthy that some works~\cite{song_denoising_2022,lu2023dpmsolver} considered the deterministic sampling process as the discretization of a continuous-time probability flow ODE. This ODE-update rule can be reversed to give a deterministic connection between $\mathbf{x}_0$ and its latent state $\mathbf{x}_t$~\cite{ho_denoising_2020}, given by
\begin{equation}
    \frac{\mathbf{x}_{t+1}}{\sqrt{\beta_{t+1}}} - \frac{\mathbf{x}_t}{\sqrt{\beta_t}} = \left( \sqrt{\frac{1 - \beta_{t+1}}{\beta_{t+1}}} - \sqrt{\frac{1 - \beta_t}{\beta_t}} \right) \boldsymbol{\epsilon}_\theta^{(t)} (\mathbf{x}_t).
\end{equation}

For inference effiency, Salimans and \citet{salimans2022progressive} defined velocity $\mathbf{v}$ as the combination of a clean sample $\mathbf{x}_0$ and noise component $\boldsymbol{\epsilon}$:
\begin{equation}
    \mathbf{v}_{\phi}=cos(\phi) \boldsymbol{\epsilon} - \sin(\phi) \mathbf{x}_0,
\end{equation}
where $\phi_t = \arctan\left(\sigma_t/\alpha_t\right)$. Therefore, the DDIM sampling process can be re-wroten by:
\begin{equation}
\mathbf{z}_{\phi_t} = \cos(\phi_t) \mathbf{x}_0{\boldsymbol{\epsilon}}(\mathbf{z}_{\phi_t}) + \sin(\phi_t) \hat{\boldsymbol{\epsilon}}(\mathbf{z}_{\phi_t}),
\end{equation}
where $\hat{\boldsymbol{\epsilon}}(\mathbf{z}_{\phi}) = \left( \mathbf{z}_{\phi} - \cos(\phi) \mathbf{\hat{x}}_{\theta}(\mathbf{z}_{\phi}) \right) \sin(\phi)$. By applying the trigonometric identities, the update step can be written as 
\begin{equation}
    \mathbf{z}_{\phi_t - \delta} = \cos(\delta) \mathbf{z}_{\phi_t} - \sin(\delta) \mathbf{v}_\phi(\mathbf{z}_{\phi_t}).
\end{equation}

\textbf{Classifier-free guidance (CFG)}. CFG is applied to guide the sampling process of diffusion models with an extra condition, such as text description. With CFG, a conditional and an unconditional diffusion model are jointly trained. At the inference stage, the noise prediction can be obtained from conditional and unconditional estimates by
\begin{equation}
\boldsymbol{\epsilon}_{\theta}(\mathbf{x}_t, t, \mathbf{y}) = w \boldsymbol{\epsilon}_{\theta}(\mathbf{x}_t, t, \mathbf{y}) + (1 - w) \boldsymbol{\epsilon}_{\theta}(\mathbf{x}_t, t, \boldsymbol{\varnothing}),
\end{equation}
where $w$ is the guidance scale controlling the strength of the condition signal on the generated output, and $\boldsymbol{\varnothing}$ denotes the null token.

\subsection{Training-Free Guidance Diffusion} \label{subsec:cfg}
Recently training-free guidance diffusion methods are introduced to control the generated output by interfering with the sampling process of diffusion models. In the community of images, DDIM inversion was proposed to manipulate an image by inverting it with the corresponding prompt and re-generating a new one conditioned on a reference prompt~\cite{mokady_null-text_2023}. Prompt-to-prompt framework~\cite{hertz_prompt--prompt_2022} was introduced to edit images by adjusting text description and attention map in cross-attention layers. \cite{mokady_null-text_2023} and \cite{huberman-spiegelglas_edit_2023} preserved in the diffusion process of source images the noise variable which is then used to adjust the noise variable of current images. \cite{mou_dragondiffusion_2024} and \cite{he_manifold_2023} designed energy functions as an extra guidance on the top of noise estimation to control the sampling process of current images. In addition, some works~\cite{mou_dragondiffusion_2024,chung_style_2024} attempted to preserve the detailed information in source images by substituting the key, value vectors of the current sampling process with those of the source diffusion process. Despite the leap made in the image domain, there remains a non-trivial issue underlying in the community of audio: \textit{sounds are transparent and always overlap with each other}. In this work, we are studying manipulating a sound track from sound mixtures while maintaining the rest of sound tracks in the audio.

\subsection{Existing Works on Audio Editing} \label{subsec:existing_work}
A straightforward approach for audio editing is to train a controllable audio generative model capable of taking extra conditions as guidance. AudioBox~\cite{team_audiobox_nodate}, a flow-matching model conditioned on both text and audio prompts, was proposed to create the audio content by masking and audio infilling. \citet{wang_audit_2023} and \citet{han_instructme_2023} trained dedicated diffusion models for various audio editing tasks, such as addition, removal, replacement, and remixing. While these methods can be used for audio editing, large-scale training is required for a satisfying result, which could be impractical in some scenarios.

Some recent works focused on fine-tuning off-the-shelf models for audio editing~\cite{wang_audit_2023}. \citet{lin_arrange_2024} finetuned MusicGen~\cite{copet_simple_2023} on multiple music editing tasks by introducing extra signals as guidance. \citet{plitsis_investigating_2023} investigated several image editing methods, such as DreamBooth~\cite{ruiz_dreambooth_2023} and Textual inversion~\cite{gal_image_2022}, for audio personalization. Despite the training cost is minimal, they still need to tune the model on task-specific datasets.

Zero-shot audio editing tasks were introduced by inversing the diffusion process. \citet{liu_audioldm_2023} firstly demonstrated the potential of text-to-audio diffusion models for editing tasks using DDIM inversion. More recently, \citet{manor_zero-shot_2024} applied an edit-friendly DDPM latent space to edit the audio content by word swapping. However, such methods require precise text descriptions for transcription, limiting themselves from some editing use cases. 

\section{Audio Latent Manipulation in the Morphing Cycle} \label{sec:morphing_cycle}

\subsection{Objective} \label{subsec:objective}
In this work, we will focus on denoising diffusion models where the sampling process will be manipulated with reference audio. The proposed AudioMorhix features: (1) \textbf{Tuning-free}: The AudioMorphix is a zero-shot editing method that does not require extra training to fit task-specific data; (2) \textbf{Audio-referenced}: Instead of text instruction~\cite{wang_audit_2023} which could be ambiguous in some use cases, the AudioMorphix takes an extra audio as reference for editing;  (3) \textbf{Versatile}: the AudioMorphix is a universal framework capabable of diverse editing tasks, including addition, removal, replacement, moving, time stretching, and pitch shifting; and (4) \textbf{Region-specific}: The AudioMorphix enables to edit a particular region of audio spectrogram while keeping the rest unchanged during editing.

Let $\text{Enc}(\cdot)$ be the transformation function mapping an input signal $x$ to latent state $z$ in the diffusion process. While previous methods~\cite{zhang_musicmagus_2024,chung_style_2024} directly control the trajectory of the generation process, empirically we found: 

Premise (Latent Spatial Consistency). \textit{The spatial information of $x$ can be inferred from the latent representation $z$, such that:}
\begin{equation}
    \text{sim}(\mathbf{z}_i, \mathbf{z}_j)=\text{sim}(Enc(\mathbf{x}_i), Enc(\mathbf{x}_j)) \propto \text{sim}(\mathbf{x}_i, \mathbf{x}_j)
\end{equation}

This premise is in line with the finding in~\cite{yang2024impus}. However, in contrast to visual modalities, manipulating the latent of a sound or a spectrogram is even harder: \textit{Sound tracks are always entangled with each other in a mixture}, resulting in one pixel in a T-F spectrogram-like representation is correlated to more than one sound track.


\subsection{Manipulate latent in the morphing cycle}
\label{subsec:manipulate_latent_in_morphing_cycle}

Let reference audio $\mathbf{x}_{r}$ be the interested sound and context audio $\mathbf{x}_{c}$ be the rest of the sounds in the mixture. The mixture $\mathbf{x}_{m}$ is the combination of reference sound $\mathbf{x}_{r}$ and context $\mathbf{x}_{c}$, such that $\mathbf{x}_{m}=\mathbf{x}_{r}+\mathbf{x}_{c}$. According to our observation, the latent of the mixture $\mathbf{z}_{m}$ also correlates with the latent of foreground and background sounds, $\mathbf{z}_{r}$ and $\mathbf{z}_{c}$, by: $\mathbf{z}_{m}\propto \mathbf{z}_{r}+\mathbf{z}_{c}$. 

\begin{figure*}
    \centering
    \includegraphics[width=0.6\linewidth]{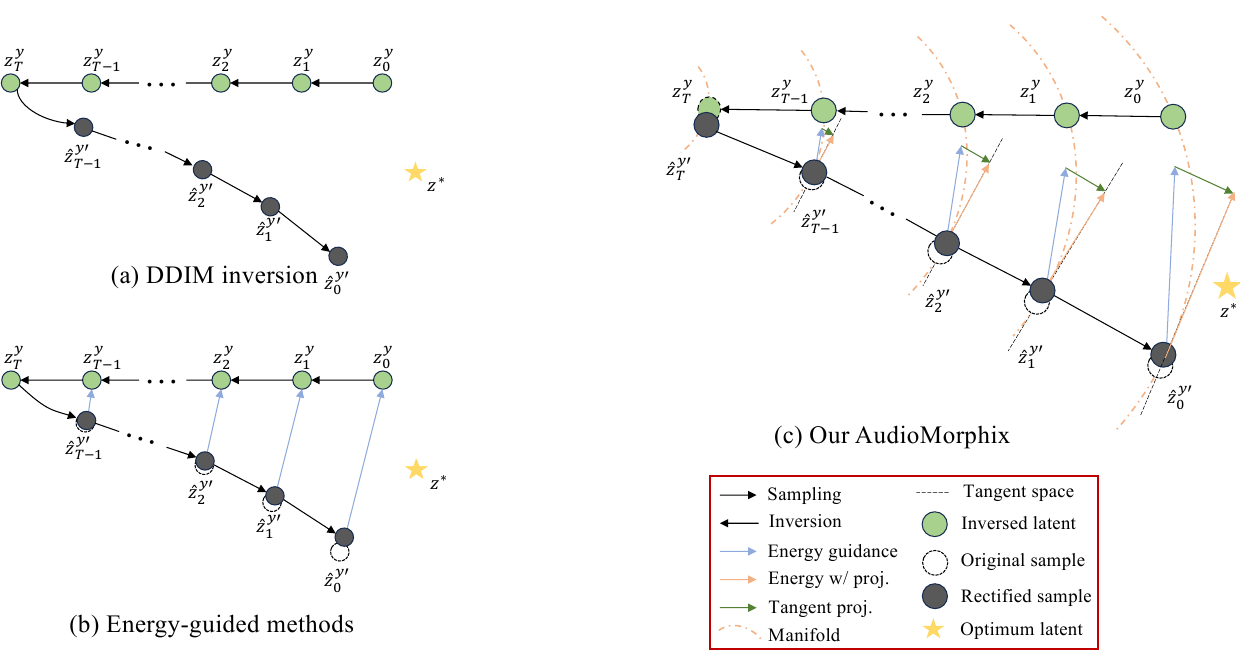}
    \caption{A schematic overview of our AudioMorphix in comparison with DDIM inversion~\protect\cite{mokady_null-text_2023} and energy-guided methods~\protect\cite{mou_dragondiffusion_2024}. Let $z_t^y$ be the latent $z$, correlated with text description $y$, at the time step $t$. We omit the process of encoding input audio $x$ into latent $z_0^y$ for simplicity. AudioMorphix refines the sampling processing by updating the noisy latent $z_T^y$ with latent optimization and performing the energy guidance at each time step.}
    \label{fig:trojactory}
    \vspace{-0.3cm}
\end{figure*}
Inspired by image morphing,
we consider a mixture be an interpolation of the reference and context sound in the morphing path and reformat three basic audio editing operations from the perspective of the morphing cycle:

\textbf{Audio addition}: 
Provided a raw audio $\mathbf{x}_{c}$ and a reference audio $\mathbf{x}_{r}$, audio addition is to obtain the interpolation of the two sounds. \citet{he_manifold_2023} and \citet{yang2024impus} argued that latent states are distributed on a manifold, suggesting the infeasibility of linearly combining two latent states. Therefore, we interpolate between the latent state $\mathbf{z}_{c}$ and $\mathbf{z}_{r}$~\footnote{latent states hereby are referred to as the noise latent at step T in the diffusion process. We ignore the subscription for simplicity.} via spherical linear interpolation (SLERP) to obtain a ``meaningful'' intermediate latent state:
\begin{equation}
   \mathbf{z}_{m}=\frac{\sin((1-\alpha)\omega)}{\sin \omega} \mathbf{z}_{c} + \frac{\sin(\alpha \omega)}{\sin \omega} \mathbf{z}_{r}, 
\end{equation}
where $\omega$ is defined by $\omega = \arccos\left(\mathbf{z}_{c} \cdot \mathbf{z}_{r}/\|\mathbf{z}_{c}\| \|\mathbf{z}_{r}\|\right)$. The denoised result $\mathbf{z}_{m}$ is then updated via the DDIM sampling by using the conditional distribution $p_{\theta}(\mathbf{x}|\mathbf{y}_{c})$.

\textbf{Audio removal}:
Audio removal is to separate a sound track $\mathbf{x}_r$ from a mixture $\mathbf{x}_m$ using audio $\mathbf{\mathbf{\Tilde{x}}}_r$ as reference. Since the orthogonal directions of reference audio $\mathbf{\mathbf{\Tilde{x}}}_r$ are not unique in a manifold, removing one sound with the reference audio \textit{only} could yield in a satisfying editing result. To this end, we resort to another sound track $\mathbf{\Tilde{x}}_c$ to regularize the sampling process of diffusion models. Algorithm \ref{alg:optimization} demonstrates how to optimize with gradient descent a latent state for the task of audio removal.

Instead of optimizing latent state $\mathbf{\Tilde{z}}_c$ and $\mathbf{\Tilde{z}}_r$ directly, the algorithm looks for the optimum direction pointing to $\mathbf{z}_c$ and $\mathbf{z}_r$. Because $\mathbf{z}_c$ and $\mathbf{z}_r$ are distributed on a sphere, we use SLERP function $g_{S}$ and geodesic distance $d_g$ to calculate the interpolation and similarity, respectively. Assuming $\mathbf{z}_c$ and $\mathbf{z}_r$ are independent from each other, we use the similarity between them as a penalty score to regularize the optimization process. We also attempt to project the optimization direction upon the sphere to ensure the updated latent states are ``meaningful'' following previous works~\cite{he_manifold_2023}. We empirically set the number of iterations $n_{iters}=100$, learning rate $lr=1e^{-4}$ and enable the use of penalty function $P$ and tangent space projection $g_{tan}$.

\textbf{Audio replacement}: Audio replacement is to replace a sound track $\mathbf{x}_{rs}$ from a mixture $\mathbf{x}_m$ with another audio $\mathbf{x}_{rt}$. We decompose the task of audio replacement by separating audio $\mathbf{x}_{rs}$ and adding audio $\mathbf{x}_{rt}$ upon the mixture $\mathbf{x}_m$. We used the same setting as audio addition and audio removal, respectively.

\section{Stepwise Guidance in Sampling Procedure} \label{sec:stepwise_guidance}

\subsection{Overview}
\label{subsec:overview}
This section introduces a stepwise guidance to control the generation procedure using the updated audio latent in Section~\ref{sec:morphing_cycle}. Motivated by previous methods~\cite{mou_dragondiffusion_2024,he_manifold_2023}, our goal is to decompose a conditional score function  $\nabla_{\mathbf{x}_t} \log p(\mathbf{x}_t | c, \mathbf{x}^r)$ into
a text-to-audio conditional score function and a differentiable term: $\nabla_{\mathbf{x}_t} \log p(\mathbf{x}_t | c, \mathbf{x}^r) = \nabla_{\mathbf{x}_t} \log p(\mathbf{x}_t | c) + \nabla_{\mathbf{x}_t} L_t(\mathbf{x}_t; \mathbf{x}^r)$. While there are some works devising energy functions for visual editing, we further improve them by considering latent in the diffusion procedure as T-F representation. Notably, our method is compatible with different prediction objective, such as epsilon~\cite{song2021denoising} and v-prediction~\cite{salimans2022progressive}.

\begin{figure}[t]
    \centering
    \includegraphics[width=0.9\columnwidth]{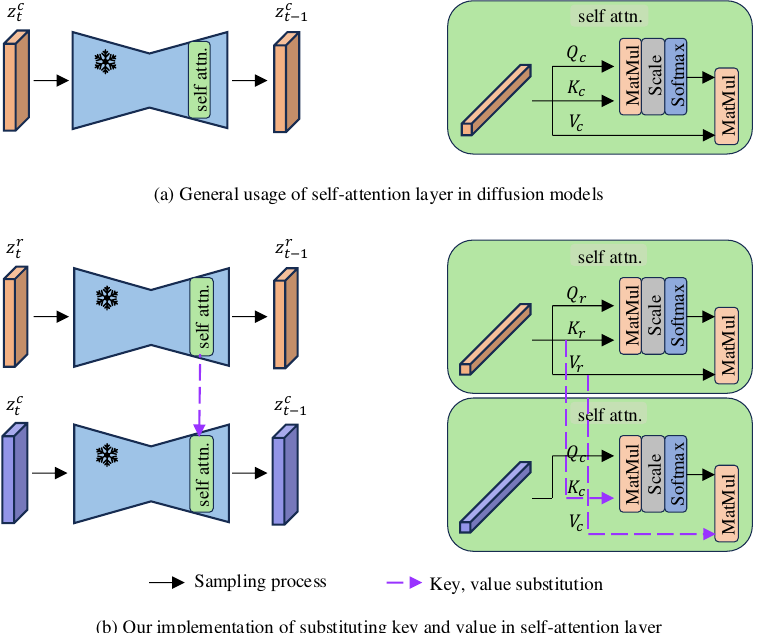}
    \caption{Illustration of adapting self-attention layers to preserve detailed information in the reference latent $\mathbf{z}_r^t$. We cache the key, value of reference latent $\mathbf{z}_r^t$ and substitute those of latent $\mathbf{z}_c^t$ during forward process.}
    \label{fig:attention}
    \vspace{-0.3cm}
\end{figure}


\begin{table*}[t]
\centering
\caption{Comparison of various audio editing methods on the AudioSet-E evaluation set, with the highest score highlighted in \textbf{bold} and the second highest in \underline{underline}.}
\label{tab:main_result}
\vspace{-0.25cm}
\begin{tabular}{@{}lcccccc@{}}
\toprule
                         & \multicolumn{2}{c}{Addition}                  & \multicolumn{2}{c}{Removal}          & \multicolumn{2}{c}{Replacement}      \\
                         & FAD~$\downarrow$ & KL~$\downarrow$ & FAD~$\downarrow$ & KL~$\downarrow$ & FAD~$\downarrow$ & KL~$\downarrow$ \\ \midrule
DDIM inversion           & \underline{5.61}          & 1.72          & 6.24          & 1.86          & 8.29          & \underline{2.05}          \\
DDPM inversion & 19.18 & 2.27 & 19.14 & 2.30 & 21.25 & 2.30 \\
AUDIT                    & 5.81          & 3.17          & \underline{3.47}          & 3.48          & \underline{5.68}          & 2.81          \\
Our method\\ (w/ AudioLDM) & \textbf{5.58} & \underline{0.83} & \textbf{2.83} & \underline{1.29} & \textbf{2.67} & 2.28 \\
Our method\\ (w/ Tango) & 6.62 & \textbf{0.57} & 6.29 & \textbf{0.77} & 7.27 & \textbf{0.62} \\
\bottomrule
\end{tabular}
\end{table*}

\begin{table*}[ht]
    \centering
    \caption{Subjective scores (\%) of various models across different tasks, with the highest score highlighted in \textbf{bold} and the second highest in \underline{underline}.}
    \label{tab:model_performance}
    \vspace{-0.25cm}
    \begin{tabular}{lccccc}
        \toprule
        & Fidelity~$\uparrow$ & Perceptual quality~$\uparrow$ & Consistency~$\uparrow$ & Region specificity~$\uparrow$ & Instruction adherence~$\uparrow$ \\
        \midrule
        Ground truth   & \textbf{60.84} & \textbf{62.27} & \textbf{60.25} & \underline{56.57} & \textbf{55.46} \\
        \midrule
        AUDIT          & 51.79 & 49.44 & 47.14 & 47.10 & 49.33 \\
        DDIM inversion & 49.32 & 50.59 & 47.63 & 49.56 & 50.08 \\
        DDPM inversion & 46.39 & 45.96 & 51.67 & 49.90 & 49.21 \\
        Our method     & \underline{56.67} & \underline{59.21} & \underline{56.73} & \textbf{58.76} & \underline{52.30} \\
        \bottomrule
    \end{tabular}
    \vspace{-0.3cm}
\end{table*}


\begin{figure*}[t]
    \centering
    \includegraphics[width=0.9\linewidth]{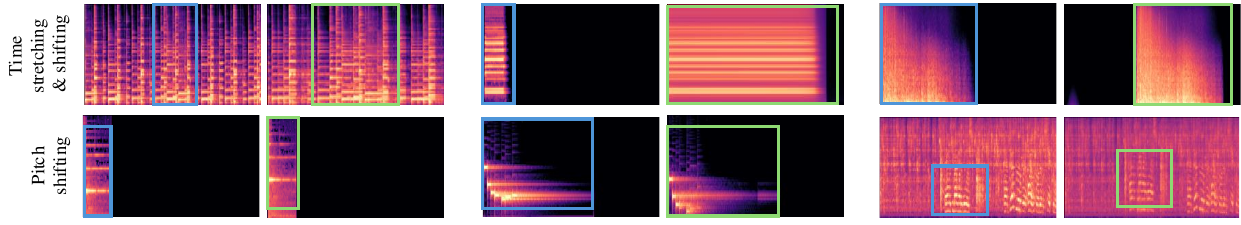}
    \vspace{-0.4cm}
    \caption{Examples of three audio manipulations: time stretching, time shifting, and pitch shifting.}
    \label{fig:tf}
\end{figure*}

\subsection{Guide Audio Editing with Energy Function}
\label{subsec:guide_audio_editing_with_energy_function}
In the AudioMorphix, various energy functions are devised as an extra guidance to control the audio generation procedure, mainly focusing on content consistency and contrast between generated audio and reference audio. 

The first derivative of an energy function is added to the score obtained from the conditional U-Net $\boldsymbol{\epsilon}_{\theta}$ for latent update in the sampling process. Suppose $\mathbf{F}_t^c$, $\mathbf{F}_t^r$ are the intermediate features obtained from the conditional U-Net $\boldsymbol{\epsilon}_{\theta}$ at step $t$ corresponding to the input audio and the reference audio, respectively. Empirically, we collate the intermediate features $\mathbf{F}_{t,l}^c$, $\mathbf{F}_{t,l}^r$ from the $l$-th self-attention layers of the U-Net decoder 
Let $\mathbf{m}^c$ and $\mathbf{m}^r$ be the binary masks upon the spectrogram of input audio and reference audio, respectively. The binary masks $\mathbf{m}^c$ and $\mathbf{m}^r$ can constrain the audio editing operating on particular T-F patches. We can measure the consistency between input and reference audio by calculating their cosine similarity over the interested area:
{\small\begin{equation}
    \text{sim}(\mathbf{F}_{t,l}^c, \mathbf{m}_c, \mathbf{F}_{t,l}^r, \mathbf{m}_r) = 0.5 \cdot \cos \left( \mathbf{F}_{t,l}^c[\mathbf{m}_c], s_g(\mathbf{F}_{t,l}^r[\mathbf{m}_r]) \right) + 0.5,
\end{equation}}
where $s_g(\cdot)$ is the gradient clipping function. Intuitively, we scale the similarity score $\text{sim}(\cdot)\in [0, 1]$ to align with human perception where $0$ means the closest distance between two audio. The guidance of consistency term is then defined by:
{\small\begin{align}
    &S_{\text{consist}}(\mathbf{F}_t^c, \mathbf{m}_c, \mathbf{F}_t^r, \mathbf{m}_r)=\nonumber \\
    &\sum_{l\in L} \frac{1}{1+4\cdot\frac{1}{HW}\sum_{h\in H}\sum_{w\in W}\text{sim}(\mathbf{F}_t^c, \mathbf{m}_c, \mathbf{F}_t^r, \mathbf{m}_r)},
\end{align}}

While the contrast concept between two audio can be defined as the reciprocal of the cosine similarity, we argue that in the audio removal use case, the sound track in the reference audio is similar but not the same as that of the input audio. Therefore, the contrast between input and reference audio is measured with the global representation of the input and the reference:
{\small\begin{equation}
    S_{\text{contrast}}(\mathbf{F}_t^c, \mathbf{m}_c, \mathbf{F}_t^r, \mathbf{m}_r) = \frac{1}{HW}\sum_{h\in H}\sum_{w\in W}\text{sim}(\mathbf{F}_t^c, \mathbf{m}_c, \mathbf{F}_t^r, \mathbf{m}_r),
\end{equation}}
Notably, the proposed energy functions are capable of generalizing to various prediction objectives, including DDIM and v-prediction, by directly modifying the probability density distribution to rectify the sampling trajectory. In experiments, we set $L=2,3$ be the selected self-attention layers of the U-Net decoder. 

\subsection{Energy guidance for Each Task}
\label{subsec:implementation}
Exploiting the consistency measurement $S_{\text{consist}}$ and contrast measurement $S_{\text{contrast}}$, we devise a variety of energy-based function:

\textbf{Audio addition}. The goal of audio addition is to mix the context audio $x^c$ with the reference audio $\mathbf{x}^r$. $\mathbf{m}_c$ and $\mathbf{m}_r$ are the binary masks of context and reference audio, respectively. Since sound tracks are ``transparent'', the original sounds in the context audio cannot be replaced with those of reference audio. Therefore, the devised energy function should consider not only the consistency between reference and generated audio, but also the consistency before and after edition. The energy-based guidance can be expressed in the following:
\begin{align}
    \boldsymbol{\epsilon}_{add}&=w_{content}\cdot S_{\text{consist}}(\mathbf{F}_t, \mathbf{m}_c,\mathbf{F}_t^c, \mathbf{m}_c)\nonumber \\
    &+w_{edit}\cdot S_{\text{consist}}(\mathbf{F}_t, \mathbf{m}_c, \mathbf{F}_t^r, \mathbf{m}_r).
\end{align}

\textbf{Audio removal}. Audio removal is to separate a sound track from the input mixture while preserving the rest of sounds. Along with pushing the generated audio away from the reference audio within the interested region of the latent space, we should also maintain the similarity of the global representation between the remaining of the original and synthesized audio:
\begin{align}
    \boldsymbol{\epsilon}_{remove}&=w_{content}\cdot S_{\text{consist}}(\mathbf{F}_t, \mathbf{m}_c,\mathbf{F}_t^c, \mathbf{m}_c)\nonumber \\
    &+w_{edit}\cdot S_{\text{contrast}}(\mathbf{F}_t, \mathbf{m}_c, \mathbf{F}_t^r, \mathbf{m}_r).
\end{align}

\textbf{Audio replacement}. We deem the replacement task as a chain of basic operations. Particularly, we exert removal and addition tasks separately to replace a sound track in the mixture with another one.

\subsection{Diffusion procedure with Memory Bank} \vspace{-0.1cm}
\label{subsec:ddim_inversion}
The combination of latent morphing and energy guidance builds a good posterior in the diffusion sampling process. However, as some works indicate, the gap between generated and reference audio still exists. Following \cite{mou_dragondiffusion_2024}, we modify the self-attention mechanism in the conditional U-Net. As shown in Figure~\ref{fig:attention}, the key, value of self-attention layers in the decoder are substituted by the original ones obtained from the inversion process. In experiments, we replace the key, value of the second and the third layers with those of the inverted trajectory. 


\begin{table*}[!ht]
\centering
\caption{Ablation study on the choices of tangent space projection and text description.}
\vspace{-0.2cm}
\label{tab:ablate}
\begin{tabular}{@{}llcccccc@{}}
\toprule
w/ Text & Tan. proj. & \multicolumn{2}{c}{Addition}      & \multicolumn{2}{c}{Removal} & \multicolumn{2}{c}{Replacement} \\
  &   & FAD~$\downarrow$ & KL~$\downarrow$ & FAD~$\downarrow$ & KL~$\downarrow$ & FAD~$\downarrow$ & KL~$\downarrow$ \\ \midrule
  &   & 6.46 & 0.84 & \textbf{2.46} & \textbf{1.03} & 6.06 & \textbf{1.00} \\
\checkmark &   &  \textbf{5.58} & \textbf{0.83} & 2.83 & 1.29 & \textbf{2.67} & 2.28    \\
  & \checkmark & 8.49 & 3.08 & 3.08  & 1.63 & 8.09 & 1.06 \\
\checkmark  & \checkmark  & 6.10 & 1.50 & 3.38 & \textbf{1.03} & 5.91 & 1.70 \\ \bottomrule
\end{tabular}
\vspace{-0.3cm}
\end{table*}

\section{Experiments} \label{sec:experiments}
\subsection{Experiment setup} \label{subsec:exp_setup} \vspace{-0.1cm}
\textbf{Datasets}. To evaluate diverse editing methods, we curated a new dataset \textit{AudioSet-E} based upon the temporally strong labeled part of AudioSet~\cite{gemmeke_audio_2017} for three audio editing tasks, including addition, removal, and replacement. AudioSet-E contains instruction, audio, and pairs of text descriptions as reference for audio editing. Particularly, AudioSet-E contains 1442 samples for audio addition, 1426 samples for audio removal, and 1870 samples for audio replacement. More about data curation in the Appendix~\ref{appendix:dataset}. 
We qualitatively evaluated the proposed AudioMorphix on moving, time stretching, and frequency shifting.

\textbf{Comparison methods}. For addition, removal, and replacement tasks, we compared our AudioMorphix against DDIM inversion~\cite{liu_audioldm_2023}, DDPM inversion~\cite{manor_zero-shot_2024}, and AUDIT~\cite{wang_audit_2023} on the AudioSet-E. We didn't implement DreamBooth and text inversion methods from~\cite{plitsis_investigating_2023} because they are targeted at audio personalization rather than manipulation. In addition to original audio, DDIM and DDPM inversion take a pair of original and target text descriptions as input while for AUDIT an editing instruction is required.

\textbf{Metrics}. We applied Frechet audio distance (FAD) and Kullback–Leibler divergence (KL) to evaluate all audio editing models. FAD measures the fidelity between generated samples and target samples while KL measures the correlation between generated samples and target samples~\citet{yuan2023leveraging}. We release our evaluation kit~\footnote{https://github.com/JinhuaLiang/TAGE} to facilitate a fair comparison in the future work.  For subjective evaluation, we assess audio editing approaches from five aspects: fidelity, perceptual quality, consistency, region specificity, and instruction adherence. The definition of these five aspects can be found in Appendix~\ref{appendix:sub_def}

\subsection{Comparisons} \label{subsec:comparison} \vspace{-0.1cm}
Table~\ref{tab:main_result} compares various audio editing methods on the AudioSet-E evaluation dataset. Our AudioMorphix outperforms the comparison methods across all tasks, particularly excelling in terms of FAD and KL metrics, which indicates better fidelity and distribution matching of the edited images. This suggests that AudioMorphix provides more accurate and realistic image edits compared to DDIM inversion, DDPM inversion, and AUDIT methods. The results are especially notable in the Addition and Removal tasks, where it shows significant improvements in the KL divergence, indicating a more precise alignment with the target distribution.

Table~\ref{tab:model_performance} shows that AudioMorphix achieves the best overall performance, outperforming all comparison methods across addition, removal, and replacement tasks. AudioMorphix surpasses the ground truth in consistency, probably because AudioMorphix adopts an end-to-end generation approach to avoid artifacts introduced by signal processing operations, such as clipping and concatenation. Additionally, the proposed model demonstrates strong perceptual coherence and localized detail preservation, making it a well-balanced generative approach. While the ground truth retains the highest fidelity score of 60.84, AudioMorphix highlights the trade-offs between strict fidelity and enhanced perceptual quality, meeting the requirement of audio editing.

Fig.~\ref{fig:tf} illustrates some examples of three audio manipulation operations, including time stretching, time shifting, and pitch shifting. Compared to traditional DSP methods that modify sounds at the waveform level, the spectrogram-based editing approach offers greater flexibility by enabling manipulation within a specified time-frequency region.


\subsection{Ablation Study} \label{subsec:ablate} \vspace{-0.1cm}
We evaluated the components of AudioMorphix by ablating text description and tangent space projection in the Table~\ref{tab:ablate}. It can be observed that the text description only significantly enhances the performance of the proposed method, achieving an FAD score of 5.58 and KL score of 0.83 on the addition task and an FAD score of 2.67 on the replacement task. Conversely, introducing tangent space projection led to performance degradation, especially on the addition and replacement tasks. This is likely because tangent space projection requires more steps for update compared to direct guidance.


\section{Conclusion} \label{sec:conclusion}
In this paper, we proposed AudioMorphix, a training-free sound editor to manipulate raw audio conditioned on reference audio and binary masks. By casting audio editing as part of a morphing cycle performed on the latent space manifold, our approach promises high-fidelity audio editing within a controllable window, while without introducing training cost, paving the way for more controllable audio editing. This approach leverages the gradient-based optimization to iteratively search the appropriate noisy latent of the current diffusion process. The proposed method incorporates various energy functions that rectify the trajectory of the sampling process. Furthermore, AudioMorphix adopts key-value substitution within self-attention layers, preserving the details of raw audio during editing. The experiments on various audio editing tasks show the effectiveness and promise of AudioMorphix compared to previous audio editing methods.


\newpage
\bibliography{src/main}
\bibliographystyle{src/named}

\clearpage
\appendix
\section*{Appendix}
\section{Dataset curation}
\label{appendix:dataset}

We curated a new dataset to evaluate various audio editing tasks, including addition, removal, and replacement, based on the temporally strong subset of the AudioSet dataset (AudioSet-SL)~\cite{gemmeke_audio_2017}. Utilizing the timestamps of sound events in AudioSet-SL, we mixed 2-3 audio tracks together with or without the selected sound events. A separate dataset was created for each task as described below:

\textbf{Audio Addition}. We randomly selected a sound event from two audio samples in the database and created two mixtures: one with and one without the selected sound event. The mixture without the selected sound event was used as the raw audio, and the mixture with the event was used as the target audio. Additionally, we used the isolated sound event as the reference audio. For text descriptions, we used a bag of sound event categories from AudioSet-SL, filling predefined templates with the selected sound event’s name as the instruction.

\textbf{Audio Removal}. The curation of the audio removal dataset follows a similar process to the audio addition task. However, the mixture with the selected sound events in the audio removal was used as the raw audio, and the mixture without those events served as the target audio. To increase the difficulty of the audio-driven editing task, we randomly sampled 1-second clips from the selected events and discarded the remaining portions during preprocessing.

\textbf{Audio Replacement}. We randomly selected three audio recordings, labeled A, B, and C, from AudioSet-SL. We ensured that A and B contained overlapping sound events from different categories. For the raw audio, we mixed audio C and the overlapped region from audio A, and for the target audio, we blended audio C and the same region from audio B. Recordings from A and B were used as the reference audio. For text descriptions, we used combinations of sound events from the two tracks (A and B), filling predefined templates with the relevant sound events as the editing instruction.

The resulting dataset, AudioSet-E, contains 1,442 samples for audio addition, 1,426 samples for audio removal, and 1,870 samples for audio replacement. Compared to previous audio editing datasets~\cite{10446663,liang2024wavcraftaudioeditinggeneration}, AudioSet-E provides a more diverse platform to evaluate the quality of generated audio across multiple editing tasks.



\section{Implement details}
\label{appendix:Implement}
We used a single NVIDIA A100 for evaluation. For a fair comparison, our AudioMorphix was provided with no masking information same as the other editing methods. We set the guidance scale to 1 for AudioLDM and 1.2 for Tango. For our AudioMorphix and ddim inversion, we set the number of inference steps as 50 while implementing DDPM inversion with 200 steps. 

\section{Subjective evaluation setup}
\label{appendix:subject_eval_setup}
Our subjective evaluations were carried out using Amazon Mechanical Turk~\footnote{https://requester.mturk.com}. We provided raters with task descriptions and detailed instructions to ensure a consistent evaluation process. Each audio sample was assessed by a minimum of 20 different raters. The final score for each system was calculated by averaging scores across all raters and audio samples.

\section{Latent optimization in the audio removal task}
\label{appendix:latent_optim}

\begin{algorithm}
\caption{Latent optimization for the removal task}
\label{alg:optimization}
\textbf{Input}:
    $\mathbf{\Tilde{z}}_c$, $\mathbf{\Tilde{z}}_r$ $\mathbf{z}_m$, $t$, $lr$, $n_{iter}$, $use\_penalty$, $use\_tangent$
\textbf{Output}: $\mathbf{\hat{z}}_c$ and $\mathbf{\hat{z}}_r$
\begin{algorithmic}[1]
\STATE Initialize $\boldsymbol{\epsilon}_c \leftarrow \mathbf{0}$, $\boldsymbol{\epsilon}_r \leftarrow \mathbf{0}$ with gradients
\STATE $optimizer \leftarrow \text{SGD}([\boldsymbol{\epsilon}_c, \boldsymbol{\epsilon}_r], lr)$

\FOR{$i = 1$ to $n_{iter}$}
    \STATE $optimizer.zero\_grad()$
    
    \STATE $\mathbf{\hat{z}}_c \leftarrow \mathbf{\Tilde{z}}_c + \boldsymbol{\epsilon}_c$
    \STATE $\mathbf{\hat{z}}_r \leftarrow \mathbf{\Tilde{z}}_r + \boldsymbol{\epsilon}_r$

    \STATE $\mathbf{\hat{z}}_m \leftarrow g_S(t, \mathbf{\hat{z}}_c, \mathbf{\hat{z}}_r)$

    \STATE $loss \leftarrow g_{tan}(\mathbf{z}_m, \mathbf{\hat{z}}_m)$

    \IF{$use\_penalty$}
        \STATE $penalty \leftarrow (\sum \mathbf{\hat{z}}_c \cdot \mathbf{\hat{z}}_r)^2$
        \STATE $loss \leftarrow loss + penalty$
    \ENDIF

    \STATE $loss.backward()$

    \IF{$use\_tangent$}
        \STATE $\boldsymbol{\epsilon}_c.grad \leftarrow g_{tan}(\boldsymbol{\epsilon}_c.grad, \mathbf{\hat{z}}_c)$
        \STATE $\boldsymbol{\epsilon}_r.grad \leftarrow g_{tan}(\boldsymbol{\epsilon}_r.grad, \mathbf{\hat{z}}_r)$
    \ENDIF

    \STATE $s_g([\boldsymbol{\epsilon}_c, \boldsymbol{\epsilon}_r])$

    \STATE $optimizer.step()$
\ENDFOR

\RETURN $\mathbf{\Tilde{z}}_c + \boldsymbol{\epsilon}_c,~\mathbf{\Tilde{z}}_r + \boldsymbol{\epsilon}_r$
\end{algorithmic}
\end{algorithm}

Algorithm~\ref{alg:optimization} performs latent optimization for a removal task by iteratively refining content and removal latent variables using gradient-based optimization. The process involves calculating a loss based on the model’s output and a target, with an optional penalty term to control the interaction between content and removal variables. Additionally, a tangent loss can be applied to the gradients of the latent variables to improve optimization. The optimization is carried out using Stochastic Gradient Descent (SGD) for a predefined number of iterations, and the final latent variables are returned after the optimization loop.

\section{Qualitative evaluation}
\label{appendix:Qualitative}
\begin{figure*}[ht]
    \centering
    \includegraphics[width=\linewidth]{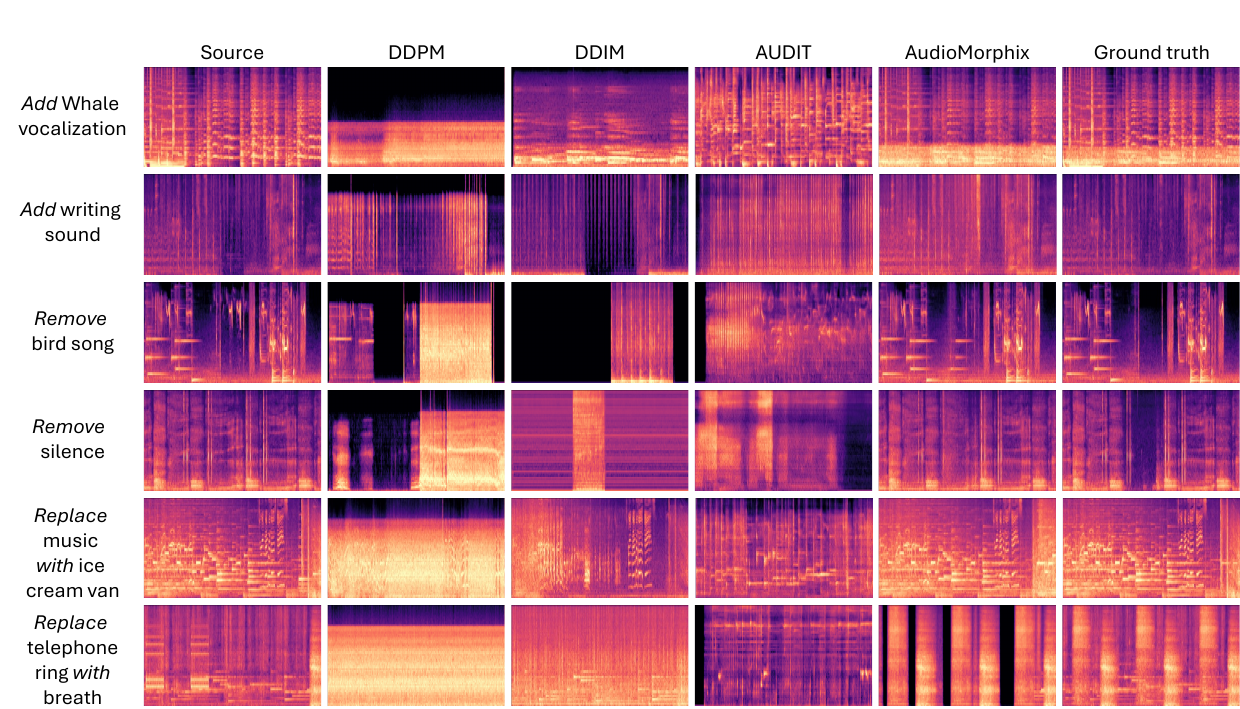}
    \caption{Qualitative evaluation between our AudioMorphix and other audio editing methods.}
    \label{fig:qual}
\end{figure*}
Figure~\ref{fig:qual} compares our proposed methods against other audio editing methods, including DDIM, DDPM, and AUDIT, over audio addition, removal, and replacement tasks. It can be observed that our AudioMorphix follows the instructions best. Additionally, the AudioMorphix remains the details of non-targeted region in the raw audio, indicating its capacity of high-fidelity audio editing.

\begin{figure*}[ht]
    \centering
    \includegraphics[width=\linewidth]{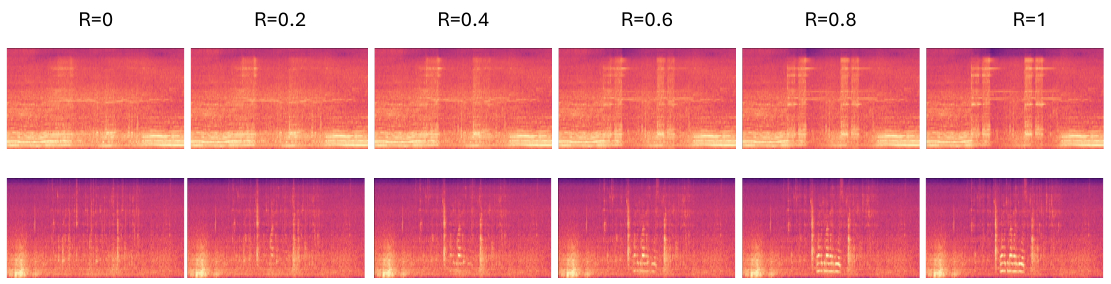}
    \caption{Ablation study on the impact of SLERP in the audio addition task.}
    \label{fig:slerp}
\end{figure*}
Figure~\ref{fig:slerp} indicates the output of the AudioMophix w.r.t. the increase of source-to-reference ratio, the ratio of source audio to the entire mixture. The goal of this experiment is to assess the impact of SLERP operations on the audio addition task. It can be observed that the generated sound smoothly morphed from the source audio to the reference audio. This supports our motivation that a sound mixture can be obtained by morphing between two different sound tracks.

\section{Definition of subjective evaluation metrics}
\label{appendix:sub_def}
To evaluate the performance of audio editing methods, we propose five distinct criteria:
\begin{itemize}
    \item Fidelity: This criterion measures how accurately the model preserves the original content, particularly in unedited portions of the audio. It ensures that no unwanted artifacts or distortions are introduced during the editing process.
    \item Perceptual quality: This evaluates the overall listening experience of the edited audio, focusing on its naturalness, clarity, and freedom from degradation. It aims to ensure that the edited audio sounds cohesive and high-quality from a human listener’s perspective.
    \item Consistency: This criterion assesses the smoothness of transitions between differently processed segments—such as between edited and unedited parts of the audio. It ensures that these transitions are seamless and imperceptible, maintaining a coherent and fluid listening experience.
    \item Region specificity: This measures whether the model restricts its editing to the designated regions without unintentionally affecting other areas of the audio. It ensures that edits are confined precisely to the intended portions.
    \item Instruction adherence: This evaluates how well the model interprets and follows the specific editing instructions provided by the user, ensuring that the modifications align with the intended changes.
\end{itemize}

\section{Subjective evaluation on different tasks}
\label{appendix:sub_eval_per_task}
\begin{figure*}[ht]
    \centering
    \includegraphics[width=\linewidth]{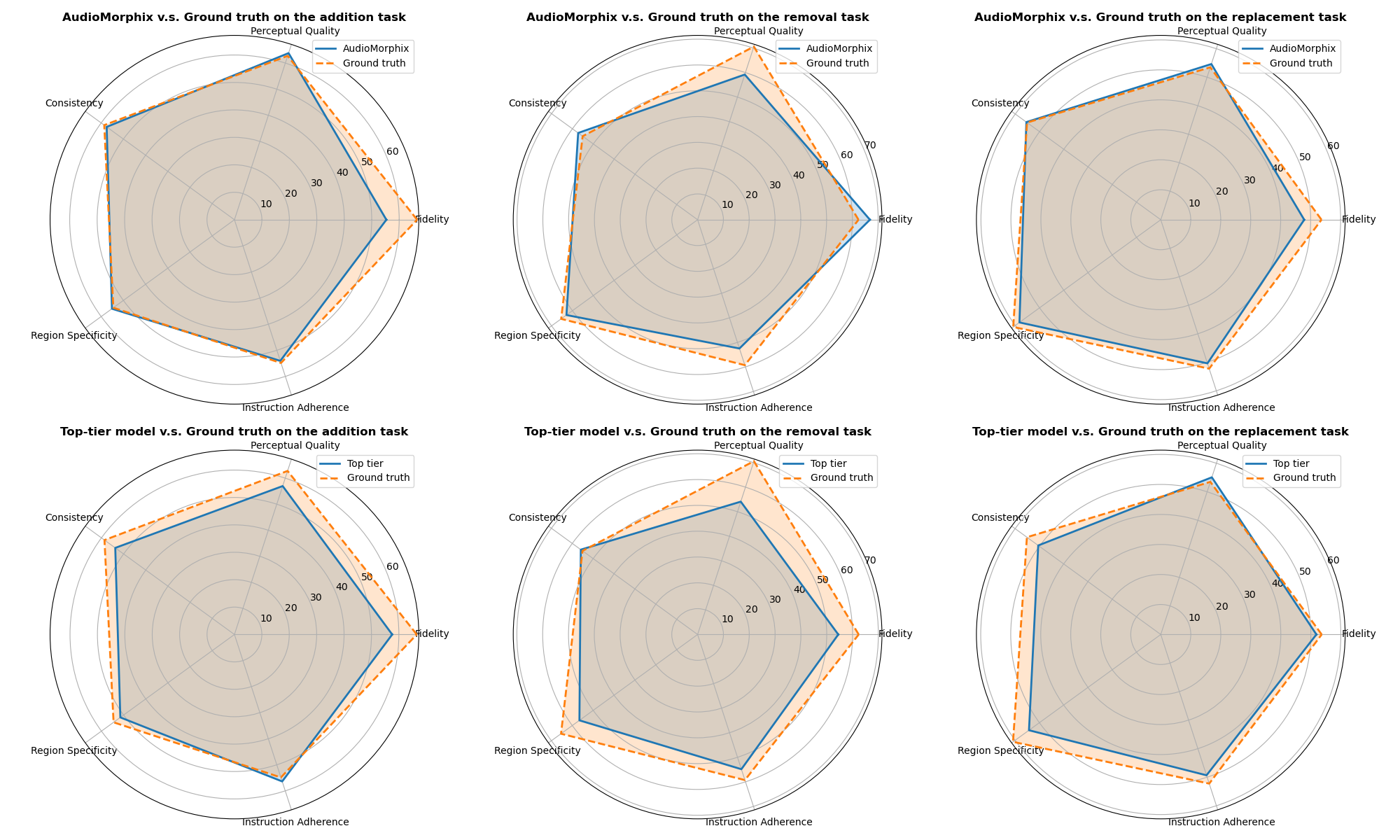}
    \caption{Subjective evaluation across different tasks, with ``top tie'' denoting the highest value among AUDIT, DDIM inversion, and DDPM inversion. }
    \label{fig:radius}
\end{figure*}
Figure~\ref{fig:radius} illustrates that AudioMorphix outperforms the top-tier model in region specificity and fidelity across various tasks, demonstrating its superior ability to edit specific regions of the target content while preserving the rest of the audio unchanged. This also suggests that AudioMorphix excels in tasks that require precise content manipulation. Furthermore, AudioMorphix shows better consistency compared to Ground Truth on the removal and replacement tasks, highlighting the advantages of its end-to-end generative approach over traditional DSP-based methods. The stable and reliable results provided by AudioMorphix emphasize its potential for high-quality, consistent audio editing.

\end{document}